\documentclass[12pt,journal,onecolumn]{IEEEtran}
\IEEEoverridecommandlockouts

\usepackage{epsfig,rotating,setspace,latexsym,amsmath,epsf,amssymb,amsfonts,bm,theorem,cite,algorithm,graphicx,epsf,authblk,epstopdf,color,algpseudocode,bbm}

\newtheorem{theorem}{Theorem}

\allowdisplaybreaks

\begin{document}

\title{Relay-Aided Secure Broadcasting for Visible Light Communications\thanks{Ahmed Arafa and H. Vincent Poor are with the Electrical Engineering Department at Princeton University, NJ 08544. Emails: {\it aarafa@princeton.edu, poor@princeton.edu}.}\thanks{Erdal Panayirci is with the Department of Electrical and Electronics Engineering, Kadir Has University, Istanbul, Turkey. Email: {\it eepanay@khas.edu.tr}.}\thanks{This research was supported in part by the U.S. National Science Foundation under Grant CNS-1702808. Erdal Panayirci has been supported by the Turkish Scientific and Research Council (TUBITAK) under the 2219 International Fellowship Program and in part by KAUST under Grant No. OSR-2016-CRG5-2958-02.}}

\author{Ahmed Arafa,~\IEEEmembership{Member,~IEEE}, Erdal Panayirci,~\IEEEmembership{Life Fellow,~IEEE}, and H. Vincent Poor,~\IEEEmembership{Fellow,~IEEE}}

\maketitle

\begin{abstract}
A visible light communication broadcast channel is considered, in which a transmitter luminaire communicates with two legitimate receivers in the presence of an external eavesdropper. A number of trusted {\it cooperative} half-duplex relay luminaires are deployed to aid with securing the transmitted data. Transmitters are equipped with single light fixtures, containing multiple light emitting diodes, and receiving nodes are equipped with single photo-detectors, rendering the considered setting as a single-input single-output system. Transmission is amplitude-constrained to maintain operation within the light emitting diodes' dynamic range. Achievable secrecy rate regions are derived under such amplitude constraints for this multi-receiver wiretap channel, first for direct transmission without the relays, and then for multiple relaying schemes: {\it cooperative jamming}, {\it decode-and-forward}, and {\it amplify-and-forward}. Superposition coding with uniform signaling is used at the transmitter and the relays. Further, for each relaying scheme, {\it secure beamforming} vectors are carefully designed at the relay nodes in order to hurt the eavesdropper and/or benefit the legitimate receivers. Superiority of the proposed relaying schemes, with secure beamforming, is shown over direct transmission. It is also shown that the best relaying scheme depends on how far the eavesdropper is located from the transmitter and the relays, the number of relays, and their geometric layout.
\end{abstract}

\section{Introduction}

Visible light communications (VLC) technology is a promising candidate for future high-speed indoor communication systems, offering solutions to spectrum congestion issues in conventional radio frequency (RF) systems \cite{komine-nakagawa-vlc, grubor-vlc}. The broadcast property in VLC, however, calls for careful design of secure communications to protect legitimate users from potential eavesdroppers, especially in public areas. Physical layer security is a powerful technique to deliver provably secure data for wireless systems through jointly encoding for reliability and security, see, e.g., \cite{poor-wireless-pls}. In this work, we design physical layer secure relaying schemes for a broadcast VLC channel with an external eavesdropper.

Recently, there have been several works on physical layer security aspects in VLC, see, e.g., \cite{mostafa-vlc-pls-jam, alouini-vlc-jam, mostafa-vlc-pls-miso, mostafa-vlc-pls-robust, arfaoui-vlc-pls-miso, arfaoui-vlc-pls-distribution, arfaoui-vlc-pls-discrete, arfaoui-vlc-pls-closed-discrete, arfaoui-vlc-pls-mimo, ding-vlc-pls-spatial, cho-vlc-pls-random, cho-vlc-pls-collude, marzban-vlc-pls-hybrid, ding-vlc-pls-eh, yin-haas-vlc-pls-multiuser, cho-vlc-sec-beamforming, cho-reflection-vlc-security, pham-vlc-pls-bccm, arfaoui-vlc-pls-bccm}. The idea of employing an external friendly node that transmits jamming signals to degrade the eavesdropper channel is investigated in \cite{mostafa-vlc-pls-jam, alouini-vlc-jam}, under amplitude constraints that are imposed such that the light emitting diodes (LEDs) operate within their dynamic range, with \cite{mostafa-vlc-pls-jam} focusing on uniform signaling and \cite{alouini-vlc-jam} focusing on truncated Gaussian signaling. Achievable secrecy rates for the multiple-input single-output (MISO) VLC channel are derived in \cite{mostafa-vlc-pls-miso}, which are then used for transmit beamforming signal design for the MISO setting in \cite{mostafa-vlc-pls-robust}. References \cite{arfaoui-vlc-pls-miso, arfaoui-vlc-pls-distribution} also derive achievable secrecy rates for the MISO VLC channel and design transmit beamforming signals, yet with a focus on truncated generalized normal signaling, showing improvement over rates achieved by both uniform and truncated Gaussian signaling. Further improvements are later shown in \cite{arfaoui-vlc-pls-discrete} by using discrete signaling with finite number of mass points. Discrete signaling is also considered in \cite{arfaoui-vlc-pls-closed-discrete}, in which closed-form achievable secrecy rates for single-input single-output (SISO) VLC channels are derived. Reference \cite{arfaoui-vlc-pls-mimo} considers a multiple-input multiple-output (MIMO) VLC channel and derives achievable secrecy rates via designing transmit covariance matrices for uncorrelated symmetric logarithmic-concave input distributions. Secrecy outage probabilities are derived in \cite{ding-vlc-pls-spatial, cho-vlc-pls-random, cho-vlc-pls-collude} with multiple eavesdroppers, via tools from stochastic geometry and spatial point processes. Security aspects of hybrid VLC/RF setups are considered in \cite{marzban-vlc-pls-hybrid, ding-vlc-pls-eh}. A multiple-transmitter and multiple-eavesdropper scenario with one legitimate user is considered in \cite{yin-haas-vlc-pls-multiuser}, in which secrecy outage probabilities and ergodic secrecy rates with and without transmitters' cooperation are derived. Beamforming design techniques are proposed in \cite{cho-vlc-sec-beamforming} to provide security in cases where the locations of eavesdroppers are only statistically known. The impacts of how multipath light reflections can jeoperdize security is studied in \cite{cho-reflection-vlc-security}. References \cite{pham-vlc-pls-bccm, arfaoui-vlc-pls-bccm} are the most closely related to our work, in which broadcast VLC channels with confidential messages are considered and achievable secrecy sum rates are derived.

Motivated by their ability to improve the signal-to-noise ratio (SNR) and overall performance of optical wireless communication systems, relaying luminaires have been studied in \cite{safari_vlc_relays, alsaadi_vlc_relays, yang-vlc-relay-linear, yang-vlc-relay-tri, hussein_vlc_relays, kizilirmak-vlc-relay, uysal_vlc_relay, Zhang-hybrid-vlc-rf-relay} under various settings and assumptions, yet with no external eavesdroppers. Reference \cite{safari_vlc_relays} studies amplify-and-forward and decode-and-forward relaying schemes, and shows that multi-hop diversity gains can be provided at the destination. Both relaying schemes are also studied in \cite{alsaadi_vlc_relays} to enhance achievable rates of mobile users. References \cite{yang-vlc-relay-linear, yang-vlc-relay-tri} consider multiple relaying scenarios where ceiling lights arranged in linear and triangular topologies help each other through multi-hop transmission. Several multiple relay-assisted VLC systems are proposed in \cite{hussein_vlc_relays}, where it is shown that multi gigabit-per-second rates can be realized with simple optical modulation formats. In \cite{kizilirmak-vlc-relay}, an LED light bulb in a desk lamp is used as a relay for an OFDM-based VLC system. In \cite{uysal_vlc_relay}, a cooperative VLC system is investigated in which an intermediate light source acts as a relay terminal operating in full duplex mode. Outage probability analysis is carried out in \cite{Zhang-hybrid-vlc-rf-relay} under different relaying schemes in a hybrid VLC/RF setup.

Inspired by the above works, in this paper we investigate the role of using extra luminary sources acting as trusted {\it cooperative} half-duplex relays in securing a two-user broadcast VLC channel from an external eavesdropper. In our setting, an amplitude constraint is imposed upon the transmitted signal in order for the LEDs to operate within their dynamic range. Under such amplitude constraint, we first derive an achievable secrecy rate region, without using the relays, based on superposition coding with uniform signaling at the source. We then invoke the relays, and derive achievable secrecy rate regions for several relaying schemes: {\it cooperative jamming}, {\it decode-and-forward}, and {\it amplify-and-forward}, in all of which an amplitude constraint also applies to the relays' transmissions. For each relaying scheme, we design {\it secure beamforming} signals to maximize the achievable rates under the relays' amplitude constraints. The design of the beamforming signals is based on formulating optimization problems that are inferred from the derived achievable secrecy rates. Results show the enhancement, in general, of the achievable secrecy rates using the relays, and that the best relaying scheme highly depends on the eavesdropper's distance from the transmitter and the relays, and also on the number of relays and how they are geometrically laid out.

We note that while the methodologies involved in this work have been previously introduced for RF communications, there exists some differences that need to be carefully considered when employing them in the context of VLC. First, and as mentioned above, a physical amplitude constraint applies to all transmitted signals from the LEDs. Invoking amplitude constraints calls for new transmission signaling design. For instance, Gaussian signaling, which is optimal for additive white Gaussian noise channels with {\it average} power constraints, is not even feasible here. We work with uniform signaling, as done in some works in the VLC literature, e.g., \cite{mostafa-vlc-pls-jam, mostafa-vlc-pls-miso, mostafa-vlc-pls-robust}, and derive achievable secrecy rate regions based on superposition coding using information-theoretic tools. To the best of our knowledge, this is the first time that achievable information-theoretic secrecy rate regions are derived under amplitude constraints for multiuser VLC with cooperative relays. Our analysis yields closed-form expressions that enable optimal design of the relay beamforming vectors using linear-algebraic and optimization tools, which are shown to boost the achievable secrecy rate regions in general. Second, the VLC channel model is also different from conventional RF channel models; the indoor line-of-sight model used is largely deterministic, and strongly related to the Euclidian distance of the transmission link. The channel gain is real-valued, positive, and depends mainly on the relative locations between the nodes, in addition to some physical characteristics of the illuminating LEDs.

It is worth mentioning that sending information simultaneously to multiple users over the same resource block using superposition coding is commonly referred to, in the wireless communications literature, by non-orthogonal multiple access (NOMA) \cite{poor-noma-intro, mojtaba-multiple-access-5g}. Our approach can then be viewed as providing security at the physical layer using cooperative relays in a VLC channel in which NOMA techniques are employed.

\section{System Model}

We consider an indoor VLC channel in which a transmitter (source) communicates with two legitimate receivers (users) in the presence of an external eavesdropper. The source is mounted on the ceiling, and is equipped with one light fixture that contains multiple LEDs modulated by the same electric current signal. The two users, and the eavesdropper, are assumed to lie geometrically on a two-dimensional plane close to the floor, and are each equipped with a single photo detector (PD).

The source's LEDs are driven by a fixed, positive bias electric current that sets the illumination intensity. The data signal, $x\in\mathbb{R}$, is superimposed on the bias current to modulate the instantaneous optical power emitted from the LEDs. The source employs superposition coding \cite{cover} to transmit two messages $x_1$ and $x_2$ to the first and the second user, respectively, by setting
\begin{align}
x=\alpha x_1+(1-\alpha)x_2
\end{align}
for some $\alpha\in[0,1]$ that determines the priority of each user. In VLC, since the signal is modulated onto the intensity of the emitted light, it must satisfy amplitude (or equivalently {\it peak} power) constraints that are imposed by the dynamic range of typical LEDs to maintain linear current-light conversion and avoid clipping distortion. An amplitude constraint, $A>0$, is enforced as follows:
\begin{align} \label{eq_amp_constraint}
\alpha|x_1|+(1-\alpha)|x_2|\leq A \quad \text{a.s.}
\end{align}
The VLC channel gain between the transmitting LEDs of a light fixture and a PD is given by \cite{owc-book}
\begin{align} \label{eq_vlc_channel}
\frac{A_{det}(m+1)}{2\pi l^2}\left(\frac{|z_{diff}|}{l}\right)^{m+1},
\end{align} 
where $A_{det}$ ie the PD's physical area in squared meters, $m=-\log(2)/\log\left(\cos\phi_{\frac{1}{2}}\right)$ is the order of Lambertian emission, with $\phi_{\frac{1}{2}}$ denoting the LED semi-angle at half power, $l$ denoting the distance between the LEDs and the PD, and $z_{diff}$ denoting the vertical distance between them. Note the VLC channel gain in the above model is positive and real-valued.

Let $h_1$, $h_2$, and $h_e$ denote the channel gains between the source and the first user, second user, and eavesdropper, respectively. Without loss of generality, let $h_1>h_2$, and hence the first user decodes the second user's message first then uses successive interference cancellation to decode its own message, while the second user decodes its message by treating the first user's interfering signal as noise \cite{cover}. From this point on, we denote the first user and the second user as the {\it strong} user and the {\it weak} user, respectively. We denote by $y_1$, $y_2$, and $y_e$ the received signals, in the electric domain, at the strong user, the weak user, and the eavesdropper, respectively. These are
\begin{align}
y_1=&h_1x+n_1, \\
y_2=&h_2x+n_2, \\
y_e=&h_ex+n_e,
\end{align}
where $n_1$, $n_2$, and $n_e$ are i.i.d. $\sim\mathcal{N}(0,1)$ noise terms\footnote{We choose to normalize the noise variances in this paper for simplicity of presentation, and take that effect on the SNR into the amplitude constraint's value. That is, the SNR is now given by the square of the channel gain multiplied by the square of the amplitude constraint.}.

\begin{figure}[t]
\center
\includegraphics[scale=1]{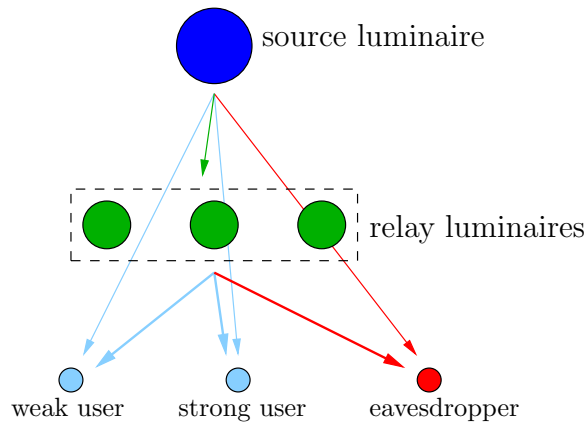}
\caption{An indoor VLC system model in which a source luminaire communicates with two legitimate users in the presence of an eavesdropper. A number of cooperative trusted relaying luminaires assist with the source's transmission.}
\label{fig_vlc_sys_mod_2}
\end{figure}

\begin{figure*}[t]
\center
\includegraphics[scale=.8]{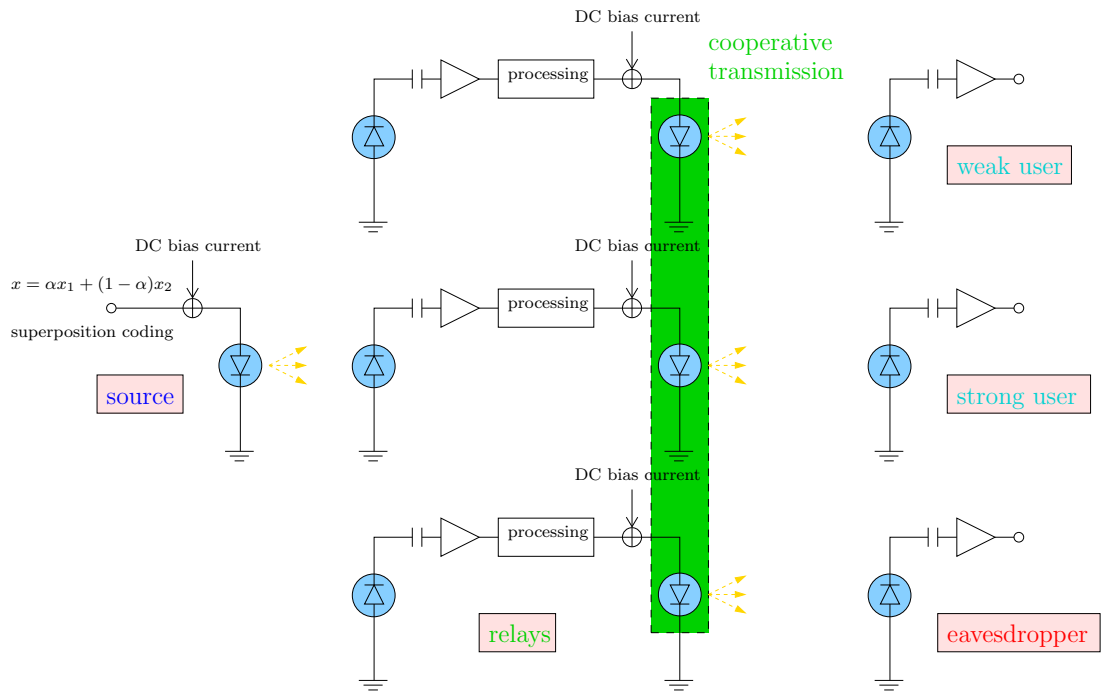}
\caption{A schematic diagram illustrating the considered VLC system model.}
\label{fig_system_schematic}
\end{figure*}

A number of extra luminary sources acting as {\it trusted cooperative} half duplex relay nodes are available to aid with securing data from the eavesdropper. Such relay nodes can be, e.g., mounted on the walls of the room in between the source and the users, or hanging from the ceiling in between them, which is possibly deployable in buildings with multi-layered lighting structures, see Fig.~\ref{fig_vlc_sys_mod_2} and the schematic diagram in Fig.~\ref{fig_system_schematic}. Let there be $K$ relays, and denote the channel gains from the source to the relays by the vector\footnote{All vectors in this paper are column vectors.} ${\bm h}_r\triangleq[h_{r,1},\dots,h_{r,K}]$. Let ${\bm g}_1$, ${\bm g}_2$, and ${\bm g}_e$ denote the $K$-length channel gain vectors from the relays to the strong user, the weak user, and the eavesdropper, respectively. All channel gains: $h_j$, $j=1,2,e$, ${\bm h}_r$ and ${\bm g}_j$, $j=1,2,e$, are assumed to be known at the source. On the other hand, the channel gains ${\bm g}_j$, $j=1,2,e$, are assumed to be known at the relays.

Similar to the source, we assume an amplitude constraint, $\bar{A}>0$, applies to the relays' transmitted signal. In order to apply a fair comparison between the relaying and non-relaying scenarios, we set $\bar{A}=\gamma A$ for some fraction $\gamma\in[0,1]$ to be designed. Operationally, we interpret the amplitude constraint as a {\it peak} power constraint at the LEDs. Hence, in case of relaying, the effective amplitude constraint that applies at the source's LEDs reduces to $A_\gamma\triangleq\sqrt{1-\gamma^2}A$. The fraction $\gamma^2$ therefore divides the total system's peak power budget $A^2$ among the source and the relays. This condition is to serve the purpose of avoiding situations in which one can add relaying LEDs at no extra cost.

In the following sections, we derive achievable secrecy rates when the source and the relays transmit their data using uniform signaling schemes. We first compute the rates without using the relays, i.e., with $\gamma=0$, and then compare them to the rates achieved under various relaying strategies: {\it cooperative jamming}, {\it decode-and-forward}, and {\it amplify-and-forward}. For these relaying schemes, we state the results for general $\gamma\in[0,1]$, and then discuss the optimal design of $\gamma$ in Section~\ref{sec_num}.

\section{Direct Transmission} \label{sec_direct}

In this section, we derive an achievable secrecy rate region via direct transmission, i.e., without using the relay nodes. We state the result in the following theorem, whose proof is in Appendix~\ref{apndx_thm_direct}:

\begin{theorem} \label{thm_direct}
The following secrecy rate pair, for the strong and weak users, is achievable via direct transmission for a given $\alpha$:\footnote{The $\log$ terms in this paper denote natural logarithms, unless explicitly specifying otherwise.}
\begin{align}
r_{1,s}=&\left[\frac{1}{2}\log\left(1+\frac{2h_1^2\alpha^2A^2}{\pi e}\right) - \frac{1}{2}\log\left(1+\frac{h_e^2\alpha^2A^2}{3}\right)\right]^+, \label{eq_r1s_direct} \\
r_{2,s}=&\left[\frac{1}{2}\log\left(\frac{1+\frac{2h_2^2A^2}{\pi e}}{1+\frac{h_2^2\alpha^2A^2}{3}}\right) - \frac{1}{2}\log\left(\frac{1+\frac{h_e^2A^2}{3}}{1+\frac{2h_e^2\alpha^2A^2}{\pi e}}\right)\right]^+, \label{eq_r2s_direct}
\end{align}
where the second subscript $s$ is to denote secrecy rates, and $[\cdot]^+\triangleq\max(\cdot,0)$.
\end{theorem}

Observe that for $\alpha=1$, we obtain $r_{2,s}=0$ since $\frac{2}{\pi e}<\frac{1}{3}$, and $r_{1,s}$ coincides with the SISO achievable secrecy rate derived in \cite{mostafa-vlc-pls-miso}, since the signal now is only directed toward one user (the strong user). The opposite holds for $\alpha=0$ as well. It is also clear from (\ref{eq_r1s_direct}) and (\ref{eq_r2s_direct}) that the strong user's achievable secrecy rate is positive if and only if (iff)
\begin{align}
\frac{2}{\pi e}h_1^2>\frac{1}{3}h_e^2,
\end{align}
and that the weak user's achievable secrecy rate is positive iff
\begin{align}
\left(\frac{2}{\pi e}-\frac{\alpha^2}{3}\right)h_2^2+\left(\frac{2\alpha^2}{\pi e}-\frac{1}{3}\right)h_e^2>\left(\frac{1}{9}-\frac{4}{\pi^2e^2}\right)\alpha^2h_2^2h_e^2.
\end{align}

Thus, achieving positive secrecy rates depends on the relative channel conditions between the users and the eavesdropper as articulated by the above inequalities. In the following sections, we study how to enhance the achievable secrecy rates in Theorem~\ref{thm_direct} above by using cooperative trusted relays.

\section{Cooperative Jamming} \label{sec_coop_jam}

In this section, we discuss the cooperative jamming scheme. In such, the relays cooperatively transmit a jamming signal ${\bm J}z$, {\it simultaneously} with the source's transmission, to confuse the eavesdropper. Here, ${\bm J}\in\mathbb{R}^K$ is a beamforming vector and $z$ is a random variable that are both to be designed under the following constraints:
\begin{align}
|z|\leq&\bar{A} \quad \text{a.s.}, \\
\|{\bm J}\|_1\leq&1,
\end{align}
where $\|\cdot\|_1$ denotes the $L_1$ norm operator: $\|{\bm J}\|_1=\sum_{i=1}^K|J_i|$. Observe that applying an $L_1$ norm constraint has the operational meaning that the cooperative relaying LEDs share the peak power budget $\bar{A}^2=\gamma^2A^2$ allocated to them, whereas if an $L_\infty$ norm is used instead, i.e., if we set: $\max_iJ_i\leq1$, then this would mean that each relay comes with its own power budget independently, i.e., the peak power budget $\bar{A}^2=\gamma^2A^2$ would be given to {\it each} relay, which would not be fair to compare with the non-relaying scenario. The received signals at the legitimate users and the eavesdropper are now given by
\begin{align}
y_1=&h_1x+{\bm g}_1^T{\bm J}z+n_1, \\
y_2=&h_2x+{\bm g}_2^T{\bm J}z+n_2, \\
y_e=&h_ex+{\bm g}_e^T{\bm J}z+n_e,
\end{align}
where the superscript $T$ denotes the transpose operation, and the amplitude constraint on the transmitted signal $x$ is now reduced to $A_\gamma=\sqrt{1-\gamma^2}A$.

In order not to harm the legitimate users, the beamforming vector is designed such that
\begin{align} \label{eq_jam_null}
{\bm g}_1^T{\bm J}={\bm g}_2^T{\bm J}=0,
\end{align}
which is guaranteed if $K\geq3$ relays, making the matrix ${\bm G}^T\triangleq[{\bm g}_1~{\bm g}_2]^T$ have a non-empty null space. Let us denote the beamforming vector satisfying (\ref{eq_jam_null}) by ${\bm J}_o$. We now have the following result, whose proof is in Appendix~\ref{apndx_thm_jam}:

\begin{theorem} \label{thm_jam}
The following secrecy rate pair, for the strong and weak users, is achievable via cooperative jamming for a given $\alpha$:
\begin{align}
r_{1,s}^J=&\left[\frac{1}{2}\log\left(1+\frac{2h_1^2\alpha^2A_\gamma^2}{\pi e}\right)- \frac{1}{2}\log\left(\frac{1+\frac{h_e^2\alpha^2A_\gamma^2}{3}+\frac{\left({\bm g}_e^T{\bm J}_o\right)^2\bar{A}^2}{3}}{1+\frac{2\left({\bm g}_e^T{\bm J}_o\right)^2\bar{A}^2}{\pi e}}\right)\right]^+, \label{eq_r1s_jam} \\
r_{2,s}^J=&\left[\frac{1}{2}\log\left(\frac{1+\frac{2h_2^2A_\gamma^2}{\pi e}}{1+\frac{h_2^2\alpha^2A_\gamma^2}{3}}\right)-\frac{1}{2}\log\left(\frac{1+\frac{h_e^2A_\gamma^2}{3}+\frac{\left({\bm g}_e^T{\bm J}_o\right)^2\bar{A}^2}{3}}{1+\frac{2h_e^2\alpha^2A_\gamma^2}{\pi e}+\frac{2\left({\bm g}_e^T{\bm J}_o\right)^2\bar{A}^2}{\pi e}}\right)\right]^+, \label{eq_r2s_jam}
\end{align}
where the superscript $J$ is to denote the cooperative jamming scheme.
\end{theorem}

We now proceed to find the optimal beamforming vector ${\bm J}_o$ that maximally degrades the eavesdropper's channel. In view of (\ref{eq_r1s_jam}) and (\ref{eq_r2s_jam}), by direct first derivative analysis, one can show that $r_{1,s}^J$ is increasing in $\left({\bm g}_e^T{\bm J}_o\right)^2$ iff
\begin{align} \label{eq_jam_inc_1}
h_e^2\alpha^2A_\gamma^2>\frac{\pi e}{2}-3\approx1.27,
\end{align}
and that $r_{2,s}^J$ is increasing in $\left({\bm g}_e^T{\bm J}_o\right)^2$ iff
\begin{align} \label{eq_jam_inc_2}
h_e^2\left(1-\alpha^2\right)A_\gamma^2>\frac{\pi e}{2}-3\approx1.27.
\end{align}
We note that, as a direct consequence of the data processing inequality \cite{cover}, sending a jamming signal can only degrade the eavesdropper's channel. It is clear, however, that the inequalities in (\ref{eq_jam_inc_1}) and (\ref{eq_jam_inc_2}) do not hold all the time, and hence sending a jamming signal might actually benefit the eavesdropper. This is justified though, since we only derive lower bounds on the achievable secrecy rates, as opposed to exact computations. Whenever the secrecy rate (of either user) is increasing in $\left({\bm g}_e^T{\bm J}_o\right)^2$, we find the optimal beamforming vector ${\bm J}_o^*$ by solving the following optimization problem:
\begin{align} \label{opt_jam_both}
\max_{{\bm J}_o}\quad&\left({\bm g}_e^T{\bm J}_o\right)^2 \nonumber \\
\mbox{s.t.}\quad&{\bm G}^T{\bm J}_o=\begin{bmatrix}0&0\end{bmatrix} \nonumber \\
&\|{\bm J}_o\|\leq 1.
\end{align}

To solve the above problem, we first introduce the following orthogonal projection notation onto the null space of ${\bm G}^T$:
\begin{align} \label{eq_p_perp}
\mathcal{P}^\perp\!\left({\bm G}\right)\triangleq{\bm I}_K-{\bm G}\left({\bm G}^T{\bm G}\right)^{-1}{\bm G}^T,
\end{align}
where ${\bm I}_K$ denotes the $K\times K$ identity matrix\footnote{Note that $\mathcal{P}^\perp\!(\cdot)$ can be defined to operate on vectors as well, denoting a projection onto their orthogonal complements in the space.}. It is clear that any vector lying in the null space of ${\bm G}^T$ can be written as the multiplication of $\mathcal{P}^\perp\!\left({\bm G}\right)$ by some vector ${\bm u}_J\in\mathbb{R}^K$. The optimal ${\bm J}_o^*$ vector then should be then of the form
\begin{align}
{\bm J}_o^*=\mathcal{P}^\perp\!\left({\bm G}\right){\bm u}_J,
\end{align}
whence the objective function of problem (\ref{opt_jam_both}) would be given by $\left({\bm g}_e^T\mathcal{P}^\perp\!\left({\bm G}\right){\bm u}_J\right)^2$, which is maximized by choosing ${\bm u}_J=c_J\mathcal{P}^\perp\!\left({\bm G}\right){\bm g}_e$, for some constant $c_J\in\mathbb{R}$. Finally, to satisfy the amplitude constraint, we choose the constant $c_J$ such that
\begin{align}
{\bm J}_o^*=\frac{\mathcal{P}^\perp\!\left({\bm G}\right){\bm g}_e}{\left\|\mathcal{P}^\perp\!\left({\bm G}\right){\bm g}_e\right\|_1}.
\end{align}

\section{Decode-and-Forward} \label{sec_dec_fwd}

In this section, we discuss the decode-and-forward scheme. Communication occurs over two phases. In the first phase, the source broadcasts its messages to both the legitimate users and relays. In the second phase, the relays decode the received messages and forward them to the users. The eavesdropper overhears the transmission over the two phases.

The received signal at the relays in the first phase is
\begin{align}
{\bm y}_r={\bm h}_rx+{\bm n}_r,
\end{align}
where ${\bm n}_r\sim\mathcal{N}({\bm 0},{\bm I}_K)$ represents the Gaussian noise in the source-relays channels. In the second phase, the $i$th relay decodes its received signal to find $x_1$ and $x_2$, re-encodes $x_1$ into $\tilde{x}_1$ and $x_2$ into $\tilde{x}_2$ using independent codewords, and then forwards them to the users using superposition coding after multiplying its transmitted signal by a constant $d_i\in\mathbb{R}$ to be designed. Effectively, the relays' transmitted signal in the second phase is given by ${\bm d}x_r$, with ${\bm d}=[d_1,d_2,\dots,d_K]$, and $x_r$ given by
\begin{align}
x_r=\alpha\tilde{x}_1+(1-\alpha)\tilde{x}_2.
\end{align}
That is, we assume the relays use the same $\alpha$ fraction as the source. The following constraints hold at the relays:
\begin{align}
\alpha|\tilde{x}_1|+(1-\alpha)|\tilde{x}_2|\leq&\bar{A} \quad \text{a.s.}, \\
\|{\bm d}\|_1\leq&1.
\end{align}
The received signals at the legitimate users and the eavesdropper in the second phase are given by
\begin{align}
y_1^r=&{\bm g}_1^T{\bm d}x_r+n_1^r, \\
y_2^r=&{\bm g}_2^T{\bm d}x_r+n_2^r, \\
y_e^r=&{\bm g}_e^T{\bm d}x_r+n_e^r,
\end{align}
where the superscript $r$ is to denote signals received from the relays, and the noise terms $n_1^r$, $n_2^r$, and $n_e^r$ are i.i.d. $\sim\mathcal{N}(0,1)$.

For the number of relays $K\geq2$, we propose designing the beamforming vector ${\bm d}$ to satisfy
\begin{align}
{\bm g}_e^T{\bm d}=0
\end{align}
so that the eavesdropper does not receive any useful information in the second phase. We denote such beamforming signal by ${\bm d}_o$. If $K\geq3$, then it will hold that both ${\bm g}_1^T{\bm d}_o$ and ${\bm g}_2^T{\bm d}_o$ are non-zero a.s. We now have the following theorem, whose proof is in Appendix~\ref{apndx_thm_dec}:

\begin{theorem} \label{thm_dec}
The following secrecy rate pair, for the strong and weak users, is achievable via decode-and-forward for a given $\alpha$:
\begin{align}
r_{1,s}^{DF}=&\frac{1}{2}\left[r_1^{DF} - \frac{1}{2}\log\left(1+\frac{h_e^2\alpha^2A^2_\gamma}{3}\right)\right]^+, \label{eq_r1s_dec} \\
r_{2,s}^{DF}=&\frac{1}{2}\left[r_2^{DF} - \frac{1}{2}\log\left(\frac{1+\frac{h_e^2A^2_\gamma}{3}}{1+\frac{2h_e^2\alpha^2A^2_\gamma}{\pi e}}\right)\right]^+, \label{eq_r2s_dec}
\end{align}
where the superscript $DF$ is to denote the decode-and-forward scheme, and $r_1^{DF}$ and $r_2^{DF}$ given by (\ref{eq_rd1}) and (\ref{eq_rd2}), respectively, at the top of this page.
\end{theorem}

\begin{figure*}[t]
\begin{align}
r_1^{DF}\!=&\min\!\left\{\!\frac{1}{2}\log\!\left(1\!+\!\frac{2h_1^2\alpha^2A^2_\gamma}{\pi e}\right)\!+\!\frac{1}{2}\log\!\left(1\!+\!\frac{2\left({\bm g}_1^T{\bm d}_o\right)^2\alpha^2\bar{A}^2}{\pi e}\right),\frac{1}{2}\log\!\left(1\!+\!\min_{1\leq i\leq K}\frac{2h_{r,i}^2\alpha^2A^2_\gamma}{\pi e}\right)\!\right\} \label{eq_rd1} \\
r_2^{DF}\!=&\min\!\left\{\!\frac{1}{2}\log\left(\frac{1+\frac{2h_2^2A^2_\gamma}{\pi e}}{1+\frac{h_2^2\alpha^2A^2_\gamma}{3}}\right)+\frac{1}{2}\log\left(\frac{1+\frac{2\left({\bm g}_2^T{\bm d}_o\right)^2\bar{A}^2}{\pi e}}{1+\frac{\left({\bm g}_2^T{\bm d}_o\right)^2\alpha^2\bar{A}^2}{3}}\right),\frac{1}{2}\log\left(\min_{1\leq i\leq K}\frac{1+\frac{2h_{r,i}^2A^2_\gamma}{\pi e}}{1+\frac{h_{r,i}^2\alpha^2A^2_\gamma}{3}}\right)\!\right\} \label{eq_rd2}
\end{align}
\hrulefill
\end{figure*}

In view of (\ref{eq_rd1}) and (\ref{eq_rd2}), we see that $r_1^{DF}$ is increasing in $\left({\bm g}_1^T{\bm d}_o\right)^2$, while direct first derivative analysis shows that $r_2^{DF}$ is increasing in $\left({\bm g}_1^T{\bm d}_o\right)^2$ iff $\alpha\leq\sqrt{\frac{2/\pi e}{1/3}}\approx0.838$, yet this condition can be ignored since $r_{s,2}^{DF}$ can only be positive if $\alpha\leq0.838$. Therefore, we propose the following optimization problem to find the best beamforming vector:
\begin{align} \label{opt_dec}
\max_{{\bm d}_o}\quad&\alpha\left({\bm g}_1^T{\bm d}_o\right)^2+(1-\alpha)\left({\bm g}_2^T{\bm d}_o\right)^2 \nonumber \\
\mbox{s.t.}\quad&{\bm g}_e^T{\bm d}_o=0 \nonumber \\
&\|{\bm d}\|_1\leq1.
\end{align}
To satisfy the first constraint, the optimal ${\bm d}_o^*$ should be of the form
\begin{align}
{\bm d}_o^*=\mathcal{P}^\perp\!\left({\bm g}_e\right){\bm u}_d\triangleq {\bm F}_d{\bm u}_d
\end{align}
for some vector ${\bm u}_d\in\mathbb{R}^K$ to be designed, with $\mathcal{P}^\perp\!(\cdot)$ as defined in (\ref{eq_p_perp}). To choose the best ${\bm u}_d$, we rewrite the objective function of the above problem slightly differently as follows:
\begin{align}
{\bm u}_d^T{\bm F}_d\left(\alpha{\bm g}_1{\bm g}_1^T+(1-\alpha){\bm g}_2{\bm g}_2^T\right){\bm F}_d{\bm u}_d.
\end{align}
Therefore, the optimal ${\bm u}_d$ is given by
\begin{align}
{\bm u}_d=c_d{\bm v}_d,
\end{align}
where $c_d\in\mathbb{R}$ is a constant, and ${\bm v}_d$ is the leading eigenvector of the matrix
\begin{align}
{\bm F}_d\left(\alpha{\bm g}_1{\bm g}_1^T+(1-\alpha){\bm g}_2{\bm g}_2^T\right){\bm F}_d,
\end{align}
i.e., the eigenvector corresponding to the largest eigenvalue of the matrix. Finally, we choose $c_d$ to satisfy the amplitude constraint as follows:
\begin{align}
{\bm u}_d=\frac{{\bm v}_d}{\left\|{\bm v}_d\right\|_1}.
\end{align}

\section{Amplify-and-Forward}

In this section, we discuss the amplify-and-forward scheme. As in the decode-and-forward scheme, communication occurs over two phases. However, in the second phase, the $i$th relay merely re-sends its received signal from the first phase after multiplying (amplifying) it by a constant $a_i\in\mathbb{R}$ to be designed. Effectively, the relays' transmitted signal in the second phase is given by $\texttt{diag}\left({\bm y}_r\right){\bm a}$, where $\texttt{diag}({\bm l})$ is the diagonalization of the vector ${\bm l}$, and the following amplitude constraint holds at the relays:
\begin{align}
\left\|\texttt{diag}\left({\bm y}_r\right){\bm a}\right\|_1\leq\bar{A} \quad \text{a.s.}
\end{align}
The received signals at the legitimate users and the eavesdropper in the second phase are given by
\begin{align}
y_1^r=&{\bm g}_1^T\texttt{diag}\left({\bm y}_r\right){\bm a}+n_1^r, \\
y_2^r=&{\bm g}_2^T\texttt{diag}\left({\bm y}_r\right){\bm a}+n_2^r, \\
y_e^r=&{\bm g}_e^T\texttt{diag}\left({\bm y}_r\right){\bm a}+n_e^r.
\end{align}

As in the decode-and-forward scheme, for $K\geq2$ relays, we propose designing the beamforming vector ${\bm a}$ to satisfy
\begin{align}
{\bm g}_e^T\texttt{diag}\left({\bm h}_r\right){\bm a}=0
\end{align}
so that the eavesdropper does not receive any useful information in the second phase. We denote such beamforming signal by ${\bm a}_o$. Further, for $K\geq3$ relays, it holds that both ${\bm g}_1^T\texttt{diag}\left({\bm h}_r\right){\bm a}_o$ and ${\bm g}_2^T\texttt{diag}\left({\bm h}_r\right){\bm a}_o$ are non-zero a.s. We now have the following theorem, whose proof is in Appendix~\ref{apndx_thm_amp}:

\begin{theorem} \label{thm_amp}
The following secrecy rate pair, for the strong and weak users, is achievable via amplify-and-forward for a given $\alpha$:
\begin{align}
r_{1,s}^{AF}=&\frac{1}{2}\left[\frac{1}{2}\log\left(1+\frac{2\kappa_1^2\alpha^2A^2_\gamma}{\pi e}\right) - \frac{1}{2}\log\left(1+\frac{h_e^2\alpha^2A^2_\gamma}{3}\right)\right]^+, \label{eq_r1s_amp} \\
r_{2,s}^{AF}=&\frac{1}{2}\left[\frac{1}{2}\log\left(\frac{1+\frac{2\kappa_2^2A^2_\gamma}{\pi e}}{1+\frac{\kappa_2^2\alpha^2A^2_\gamma}{3}}\right) - \frac{1}{2}\log\left(\frac{1+\frac{h_e^2A^2_\gamma}{3}}{1+\frac{2h_e^2\alpha^2A^2_\gamma}{\pi e}}\right)\right]^+, \label{eq_r2s_amp}
\end{align}
where the superscript $AF$ is to denote the amplify-and-froward scheme, and
\begin{align}
\kappa_j^2 \triangleq h_j^2+\frac{\left({\bm g}_j^T\emph{\texttt{diag}}\left({\bm h}_r\right){\bm a}_o\right)^2}{1+\left({\bm g}_j^T{\bm a}_o\right)^2},\quad j=1,2.
\end{align}
\end{theorem}

In view of (\ref{eq_r1s_amp}) and (\ref{eq_r2s_amp}), we see that $r_{1,s}^{AF}$ is increasing in $\kappa_1^2$, while direct first derivative shows that $r_{2,s}^{AF}$ is increasing in $\kappa_2^2$ iff $\alpha\leq\sqrt{\frac{2/\pi e}{1/3}}\approx0.838$, yet again this condition can be ignored (as we did in the decode-and-forward case) since $r_{2,s}^{AF}$ can only be positive if $\alpha\leq0.838$. Therefore, we propose the following fractional optimization problem to find the best beamforming vector that maximizes the $j$th user's rate, $j=1,2$:
\begin{align} \label{opt_amp}
\max_{{\bm a}_o}\quad&\frac{\left({\bm g}_j^T\texttt{diag}\left({\bm h}_r\right){\bm a}_o\right)^2}{1+\left({\bm g}_j^T{\bm a}_o\right)^2} \nonumber \\
\mbox{s.t.}\quad&{\bm g}_e^T\texttt{diag}\left({\bm h}_r\right){\bm a}_o=0 \nonumber \\
&\left\|\texttt{diag}\left({\bm y}_r\right){\bm a}_o\right\|_1\leq\bar{A}.
\end{align}
To solve the above fractional program, we introduce the following auxiliary problem:
\begin{align} \label{opt_amp_aux}
p_j^{AF}(\lambda)\triangleq\max_{{\bm a}_o}\quad&\left({\bm g}_j^T\texttt{diag}\left({\bm h}_r\right){\bm a}_o\right)^2-\lambda\left(1+\left({\bm g}_j^T{\bm a}_o\right)^2\right)  \nonumber \\
\mbox{s.t.}\quad&{\bm g}_e^T\texttt{diag}\left({\bm h}_r\right){\bm a}_o=0 \nonumber \\
&\left\|\texttt{diag}\left({\bm y}_r\right){\bm a}_o\right\|_1\leq\bar{A}
\end{align}
for some $\lambda\geq0$. One can show the following: 1) $p_j^{AF}(\lambda)$ is decreasing in $\lambda$; and 2) the optimal solution of problem (\ref{opt_amp}) is given by $\lambda^*$ that solves $p_j^{AF}\left(\lambda^*\right)=0$ \cite{dinkelbach-fractional-prog}. Hence, one can find an upper bound on $\lambda^*$ that makes $p_j^{AF}(\lambda)<0$ and then proceed by, e.g., a bisection search, to find $\lambda^*$. Focusing on problem (\ref{opt_amp_aux}), we first note that, to satisfy the first constraint, the optimal ${\bm a}_o$ should be of the form
\begin{align}
{\bm a}_o=\mathcal{P}^\perp\!\left(\texttt{diag}\left({\bm h}_r\right){\bm g}_e\right){\bm u}_a\triangleq {\bm F}_a{\bm u}_a
\end{align}
for some vector ${\bm u}_a\in\mathbb{R}^K$ to be designed. To choose the best ${\bm u}_a$, we rewrite the objective function as
\begin{align}
{\bm u}_a^T{\bm F}_a\left(\texttt{diag}\left({\bm h}_r\right){\bm g}_j{\bm g}_j^T\texttt{diag}\left({\bm h}_r\right)-\lambda{\bm g}_j{\bm g}_j^T\right){\bm F}_a{\bm u}_a.
\end{align}
Hence, the optimal ${\bm u}_a$ is given by
\begin{align}
{\bm u}_a=c_a{\bm v}_a,
\end{align}
where $c_a\in\mathbb{R}$ is a constant, and ${\bm v}_a$ is the leading eigenvector of the matrix
\begin{align}
{\bm F}_a\left(\texttt{diag}\left({\bm h}_r\right){\bm g}_j{\bm g}_j^T\texttt{diag}\left({\bm h}_r\right)-\lambda{\bm g}_j{\bm g}_j^T\right){\bm F}_a.
\end{align}
We choose $c_a$ to satisfy the amplitude constraint as follows:
\begin{align}
{\bm u}_a=\frac{{\bm v}_a}{\left\|\texttt{diag}\left({\bm y}_r\right){\bm v}_a|\right\|_1}\bar{A}.
\end{align}
Finally, let ${\bm a}_o^{(j)}$ be the solution of problem (\ref{opt_amp}). We propose using the following beamforming vector:
\begin{align}
{\bm a}_o^*=\alpha{\bm a}_o^{(1)}+(1-\alpha){\bm a}_o^{(2)}.
\end{align}

\section{Numerical Evaluations and Discussion} \label{sec_num}

In this section, we validate our results via numerical evaluations and discuss the relative performances of the proposed schemes in this paper. We characterize the boundary of the achievable secrecy regions of the different schemes by solving the following optimization problem for a given $\mu\in[0,1]$:
\begin{align} \label{opt_region}
\max_{\alpha,\gamma}\quad&\mu r_{1,s}^\omega+(1-\mu)r_{2,s}^\omega \nonumber \\
\mbox{s.t.}\quad&0\leq\alpha\leq1,~0\leq\gamma\leq1,
\end{align}
with $\omega\in\{J,DF,AF\}$ denoting the relaying scheme, or is simply not used in the case of direct transmission. We solve the above problem numerically using, e.g., a line search algorithm. Since the feasible set is bounded, this facilitates convergence to an optimal solution. For simplicity, we set $\lambda=1$ in the $AF$ beamforming vector optimization and do not further optimize it.

\begin{figure}[t]
\center
\includegraphics[scale=1]{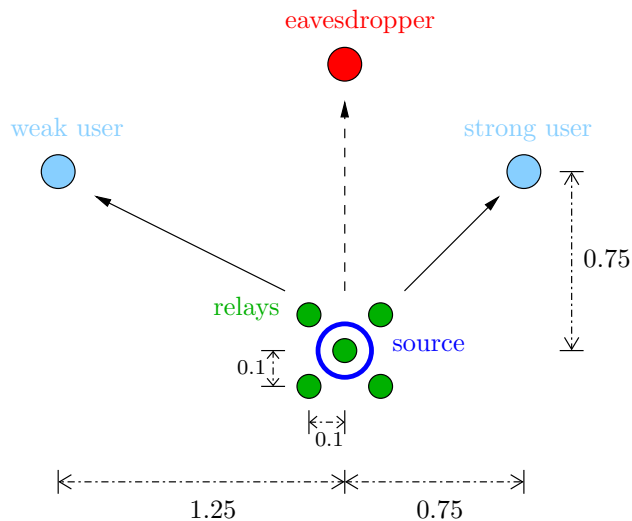}
\caption{Plan view of the geometric layout of the source, the relays, the legitimate users, and the eavesdropper.}
\label{fig_plan_view_1}
\end{figure}

We consider a room of size $5\times5\times3$ cubic meters. With the origin tuple $(0,0,0)$ denoting the center of the room's floor. The source is located at $(0,0,3)$, the strong user at $(0.75,0.75,0.7)$, and the weak user at $(-1.25,0.75,0.7)$. We consider $K=5$ relays located at the following positions: $(0.1,0.1,2)$, $(0.1,-0.1,2)$, $(0,0,2)$, $(-0.1,0.1,2)$, and $(-0.1,-0.1,2)$, see the plan view in Fig.~\ref{fig_plan_view_1}. The channel gain between two nodes is given by (\ref{eq_vlc_channel}), with $A_{det}=10^{-4}$ and $\phi_{\frac{1}{2}}=60^\circ$. We set the amplitude constraint (or the system's peak power budget) to $A=10^7$.

\begin{figure}[t]
\center
\includegraphics[scale=.5]{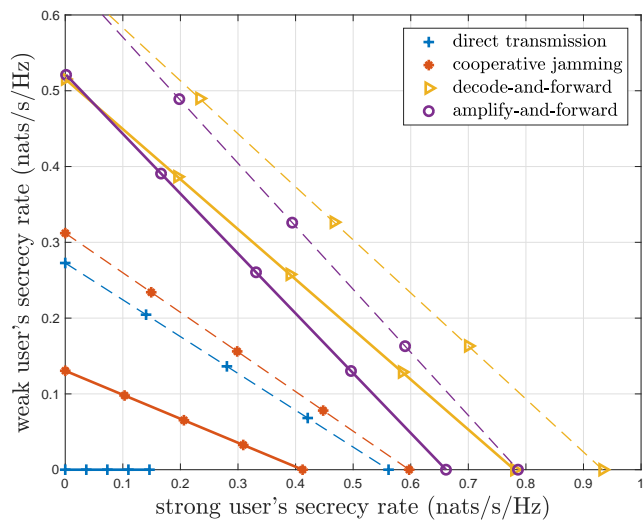}
\caption{Achievable secrecy regions of the proposed schemes. Solid lines are with eavesdropper at $(0,1.5,0.7)$, and dashed lines are with it at $(0,2,0.7)$.}
\label{fig_sec_reg}
\end{figure}

In Fig.~\ref{fig_sec_reg}, the achievable secrecy rate regions of the schemes proposed in this paper, along with that of the direct transmission scheme are shown. The solid lines in Fig.~\ref{fig_sec_reg} are when the eavesdropper is located at $(0,1.5,0.7)$. We see in this case that all the proposed schemes perform strictly better than direct transmission. The dashed lines in Fig.~\ref{fig_sec_reg} are when the eavesdropper is located a bit further away from the source (and the relays) at $(0,2,0.7)$. We see in this case that larger secrecy rates are achievable for all schemes, and that direct transmission is now comparable to cooperative jamming. We also note that they are both performing {\it closer} in this case to decode-and-forward and amplify-and-forward. The main reason behind this is that as the eavesdropper gets further away from the source, the {\it rate of increase} in the achievable secrecy rates in case of direct transmission and cooperative jamming becomes {\it larger} than that of decode-and-forward and amplify-and-forward. This is attributed to the pre-log $\frac{1}{2}$ terms in the case of decode-and-forward and amplify-and-forward that are due to the half-duplex operation of the relays. These terms have a diminishing effect on the achievable secrecy rates that becomes more apparent as the eavesdropper gets further away, whence direct transmission and cooperative jamming start performing better.

\begin{figure}[t]
\center
\includegraphics[scale=.5]{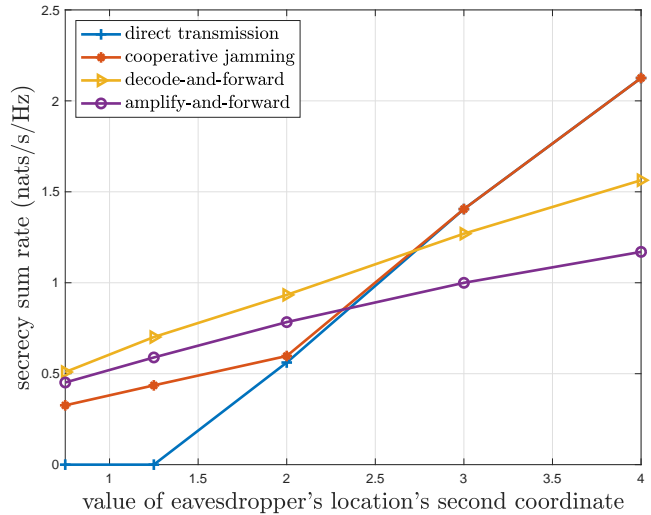}
\caption{Effect of eavesdropper's distance from the source on the achievable secrecy sum rate. Only the second coordinate of the eavesdropper's location is varied, while the first and the third coordinates are fixed at $0$ and $0.7$, respectively.}
\label{fig_sec_sum}
\end{figure}

In Fig.~\ref{fig_sec_sum}, we investigate this latter note further, and show the effect of the eavesdropper's distance from the source on the secrecy sum rate, setting $\mu=\frac{1}{2}$ in problem (\ref{opt_region}). We vary the eavesdropper's location from $(0,0.75,0.7)$ to $(0,4,0.7)$, i.e., we only change its location's second coordinate's value. We observe from the figure that clearly the secrecy sum rate increases, for all schemes, as the eavesdropper's distance from the source increases. We also note that at relatively close locations, the proposed relaying schemes achieve strictly positive rates, as opposed to the zero rate achieved via direct transmission. This shows how useful the proposed relaying schemes become, compared to direct transmission, when the eavesdropper is relatively close to the source. Finally, it can be seen from the figure that there exists a certain distance after which direct transmission and cooperative jamming beat decode-and-forward and amplify-and-forward. This is attributed to, as discussed before, the diminishing effects of the pre-log $\frac{1}{2}$ terms in the case of decode-and-forward and amplify-and-forward, which are not present in direct transmission and cooperative jamming.

\begin{figure}[t]
\center
\includegraphics[scale=.5]{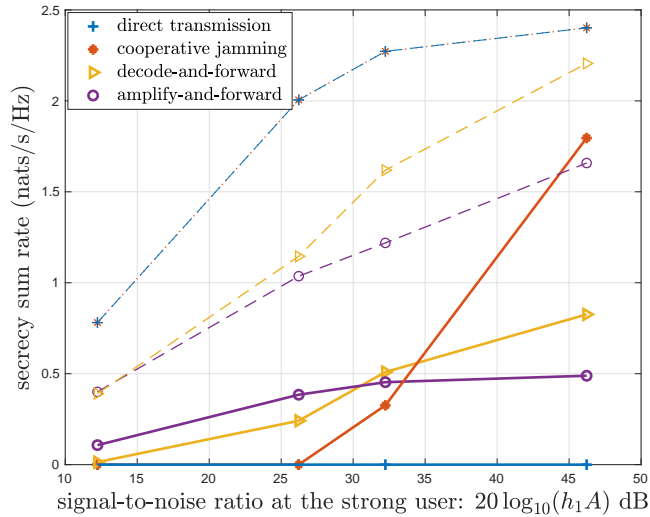}
\caption{Effect of the strong user's SNR on the achievable secrecy sum rate. Solid lines are with the eavesdropper at $(0,0.75,0.7)$, and dashed lines are with it at $(0,4.25,0.7)$.}
\label{fig_sec_sum_snr}
\end{figure}

In Fig.~\ref{fig_sec_sum_snr}, we show the effect of the SNR at the strong user on the achievable secrecy sum rates. The strong user's SNR (in dB) is given by $20\log_{10}\left(h_1A\right)$. We consider a setting in which the eavesdropper is close-by at $(0,0.75,0.7)$, whose results are depicted in solid lines, and another setting in which the eavesdropper is far-away at $(0,4.25,0.7)$, whose results are depicted in dashed lines. In the close-by setting, direct transmission achieves zero rate for all values of the SNR, cooperative jamming starts achieving positive rates only for relatively higher values of the SNR and continues to eventually beat all other schemes, amplify-and-forward performs best at relatively lower SNR values and is beaten by decode-and-forward at relatively higher ones. In the far-away setting, direct transmission and cooperative jamming are indistinguishable, and beat decode-and-forward and amplify-and-forward for all values of the SNR. This is, once more, the effect of the half-duplex operation of the relays. It is clear from Figs.~\ref{fig_sec_reg},~\ref{fig_sec_sum},~and~\ref{fig_sec_sum_snr} that the best relaying scheme depends on the secrecy rate region's operating point, the distance between the source and the eavesdropper and the SNR.

\begin{figure}[t]
\center
\includegraphics[scale=1]{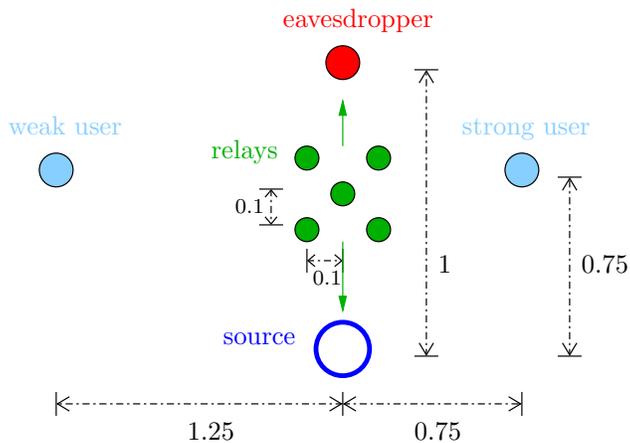}
\caption{Plan view of the geometric layout of the system, in which the center point of the relays' positions is varying.}
\label{fig_plan_view_2}
\end{figure}

\begin{figure}[t]
\center
\includegraphics[scale=.5]{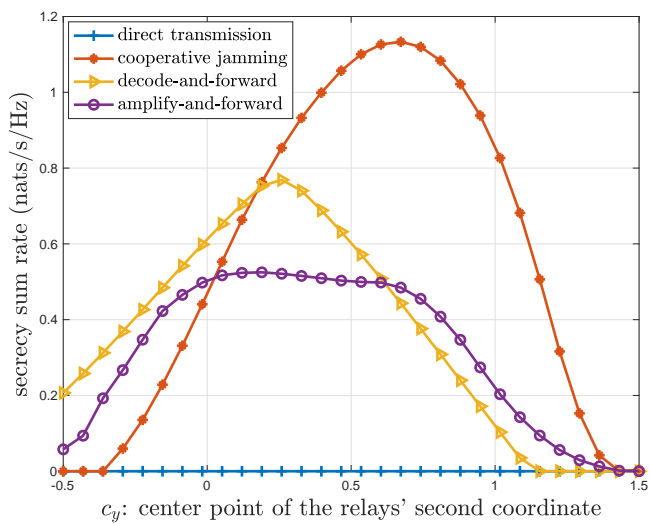}
\caption{Effect of the relays' distance from the eavesdropper on the secrecy sum rate. The eavesdropper is located at $(0,1,0.7)$, while the relays are located at $(0.1,c_y+0.1,2)$, $(0.1,c_y-0.1,2)$, $(0,c_y,2)$, $(-0.1,c_y+0.1,2)$, and $(-0.1,c_y-0.1,2)$.}
\label{fig_sum_relays_ctr}
\end{figure}

Next, we explore another aspect of relative distances between the nodes by fixing the eavesdropper's location at $(0,1,0.7)$ and varying the centroid of the relays' positions. Specifically, we let the relays be located at $(0.1,c_y+0.1,2)$, $(0.1,c_y-0.1,2)$, $(0,c_y,2)$, $(-0.1,c_y+0.1,2)$, and $(-0.1,c_y-0.1,2)$ and vary the center point $c_y$ from $-0.5$ to $1.5$, see the plan view in Fig.~\ref{fig_plan_view_2}. We plot the achievable secrecy sum rates versus $c_y$ in Fig.~\ref{fig_sum_relays_ctr}. We see from the figure that direct transmission achieves zero secrecy rates for all values of $c_y$, since the eavesdropper is relatively closer to the source than the legitimate users. On the other hand, all the proposed relaying schemes achieve strictly positive secrecy rates, with varying performances. We notice, in particular, that the relatively simple cooperative jamming scheme performs best when the relays are closest to the eavesdropper.

\begin{figure}[t]
\center
\includegraphics[scale=1]{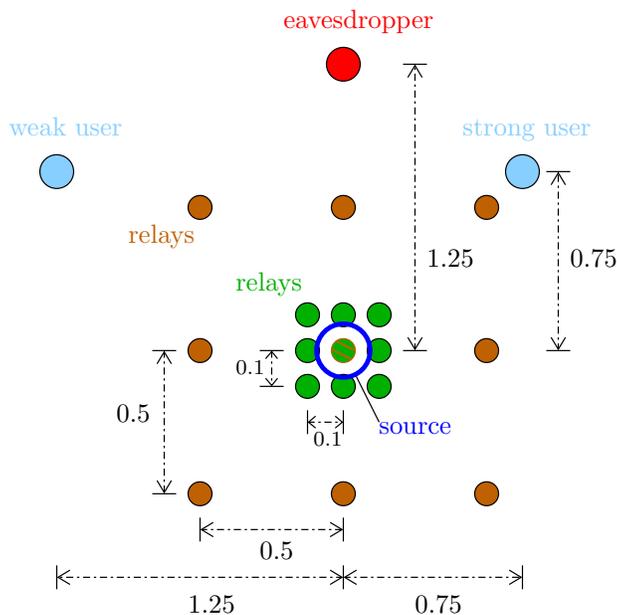}
\caption{Plan view of the geometric layout of the system, in which the number of relays is varying, as well as their relative distance from each other. Either the layout in green with $\ell=0.1$, or that in brown with $\ell=0.5$ is chosen to employ the varying number of relays.}
\label{fig_plan_view_3}
\end{figure}

\begin{figure}[t]
\center
\includegraphics[scale=.5]{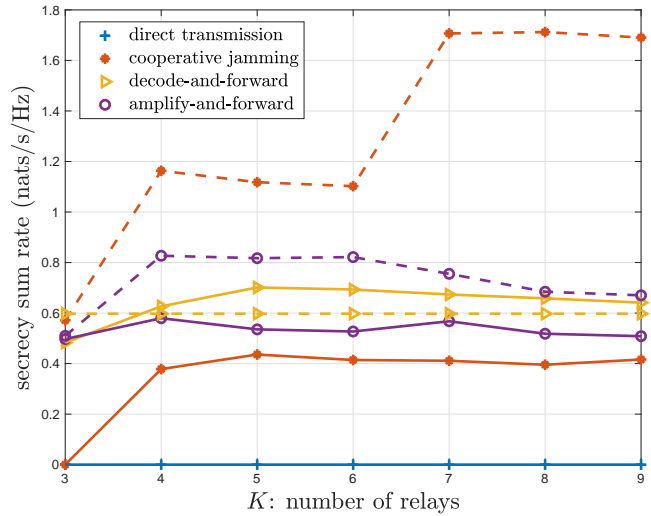}
\caption{Effect of number of relays on the secrecy sum rates of the proposed schemes. The eavesdropper is located at $(0,1.25,0.7)$. The relays are located along the corner and mid-side points of a square of side length $2\ell$ meters, centered at $(0,0,2)$. Solid lines are when $\ell=0.1$, and dashed lines are when $\ell=0.5$.}
\label{fig_sum_sec_nmbr_relays}
\end{figure}

Finally, we explore the effect of a different aspect on the secrecy sum rate: the number of relay nodes, and how far apart they are from each other. We consider the situation in which the eavesdropper is located relatively close to the source at $(0,1.25,0.7)$, and place a varying number of relays along the corners and sides of a square of side length $2\ell$ meters, centered at $(0,0,2)$. Specifically, we locate one relay at the center of the square, at $(0,0,2)$, and the remaining relays at either the corners: $(\ell,\ell,2)$, $(-\ell,\ell,2)$, $(\ell,-\ell,2)$, and $(-\ell,-\ell,2)$; or at the centers of the sides: $(\ell,0,2)$, $(0,\ell,2)$, $(-\ell,0,2)$, and $(0,-\ell,2)$, see the plan view in Fig.~\ref{fig_plan_view_3}. We vary the number of relays, $K$, from $3$ to $9$ relays, and plot the achievable secrecy sum rate for each case in Fig.~\ref{fig_sum_sec_nmbr_relays}. The solid lines in the figure are when $\ell=0.1$ meters, while the dashed lines are when $\ell=0.5$ meters. We see from the figure that direct transmission achieves zero secrecy rates, since the eavesdropper is relatively closer to the source than the legitimate users, while all the proposed schemes achieve strictly positive secrecy sum rates. The main message conveyed by this figure, however, is that for every relaying scheme, there exists an optimal number of relays that maximizes the secrecy sum rate. Such optimal number is not necessarily the maximum number of relays available ($9$ in this case). The reason behind this is that when new relay LEDs are added to the system, the power share {\it per-relay} decreases. This might hurt the overall performance if, for instance, this newly added relay is not very well-positioned with respect to the eavesdropper, relative to the already existing ones, and ends up consuming power unnecessarily. Another observation from Fig.~\ref{fig_sum_sec_nmbr_relays} is that the relative distance between the relays is an important system aspect that should be carefully designed to meet a desired system performance.

\section{Conclusion and Future Directions}

A VLC broadcast channel in which a transmitter communicates with two legitimate receivers in the presence of an external eavesdropper has been considered. Under an amplitude constraint, imposed to allow the LEDs to operate within their dynamic range, an achievable secrecy rate region has been derived, based on superposition coding with uniform signaling. Then, trusted cooperative half-duplex relay nodes have been introduced in order to assist with securing the data from the eavesdropper via multiple relaying schemes: cooperative jamming, decode-and-forward, and amplify-and-forward. Secure beamforming signals have been carefully designed at the relays to enhance the achievable secrecy rates. It has been shown that the best relaying scheme varies according to the distance from the transmitter (and the relays) to the eavesdropper, and also on the number of relays and their geometric layout.

Extending the approaches in this paper to the case with multiple transmitting LED fixtures and/or multiple receiving PDs would be of interest as a future direction. In addition, one could also consider deriving achievable secrecy rate regions based on different distributions other than uniform, such as discrete and truncated generalized normal distributions, that have been previously used in the literature. Another direction would be to consider the case in which the eavesdropper's location is not known at the transmitter, or known within some boundaries. In the former case, the goal would be deriving secrecy outage probabilities, while in the latter case, the goal could be deriving a worst case achievable secrecy rate region.

\appendix

\subsection{Proof of Theorem~\ref{thm_direct}} \label{apndx_thm_direct}

Given $\alpha$, the following secrecy rates, for the strong and weak users, are achievable for this multi-receiver wiretap channel \cite{ersen-mr-wt}:
\begin{align}
c_{1,s}=&\left[\mathbbm{I}(x;y_1|x_2)-\mathbbm{I}(x;y_e|x_2)\right]^+ \label{eq_c1s}, \\
c_{2,s}=&\left[\mathbbm{I}(x_2;y_2)-\mathbbm{I}(x_2;y_e)\right]^+, \label{eq_c2s}
\end{align}
where $\mathbbm{I}(\cdot;\cdot)$ denotes the mutual information measure \cite{cover}. Now let the transmitted symbols $x_1$ and $x_2$ represent two independent uniformly distributed random variables on $[-A,A]$. Clearly, this satisfies the amplitude constraint in (\ref{eq_amp_constraint}). Let us now drop the superscript $+$ for simplicity of presentation. We proceed by lower bounding $c_{1,s}$ as follows:
\begin{align}
c_{1,s}=&\mathbbm{I}\left(x;h_1(\alpha x_1+(1-\alpha)x_2)+n_1|x_2\right)-\mathbbm{I}\left(x;h_e(\alpha x_1+(1-\alpha)x_2)+n_e|x_2\right) \\
=&\mathbbm{I}\left(x_1;h_1\alpha x_1+n_1\right)-\mathbbm{I}\left(x_1;h_e\alpha x_1+n_e\right) \\
=&\mathbbm{h}\left(h_1\alpha x_1+n_1\right)-\mathbbm{h}\left(h_e\alpha x_1+n_e\right) \label{eq_r1s_direct_pf_1} \\
\geq&\frac{1}{2}\log\left(e^{2\mathbbm{h}\left(h_1\alpha x_1\right)}+e^{2\mathbbm{h}\left(n_1\right)}\right)-\frac{1}{2}\log\left(2\pi e\left(h_e^2\alpha^2\frac{A^2}{3}+1\right)\right) \label{eq_r1s_direct_pf_2} \\
=&\frac{1}{2}\log\left(h_1^2\alpha^24A^2+2\pi e\right)-\frac{1}{2}\log\left(2\pi e\left(h_e^2\alpha^2\frac{A^2}{3}+1\right)\right) \\
=&r_{1,s},
\end{align}
where $\mathbbm{h}(\cdot)$ in (\ref{eq_r1s_direct_pf_1}) denotes the differential entropy measure \cite{cover}, and (\ref{eq_r1s_direct_pf_2}) follows by lower bounding the first (positive) term in (\ref{eq_r1s_direct_pf_1}) by the entropy power inequality (EPI) \cite{cover} and upper bounding the second (negative) term in (\ref{eq_r1s_direct_pf_1}) by plugging in a Gaussian $x_1$, instead of uniform, with the same variance, since Gaussian maximizes differential entropy \cite{cover}. Next, we proceed similarly to lower bound $c_{2,s}$ as follows:
\begin{align}
c_{2,s}=&\mathbbm{I}\left(x_2;h_2(\alpha x_1+(1-\alpha)x_2)+n_2\right)-\mathbbm{I}\left(x_2;h_e(\alpha x_1+(1-\alpha)x_2)+n_e\right) \\
=&\mathbbm{h}\left(h_2(\alpha x_1+(1-\alpha)x_2)+n_2\right)-\mathbbm{h}\left(h_2\alpha x_1+n_2\right) \nonumber \\
&-\mathbbm{h}\left(h_e(\alpha x_1+(1-\alpha)x_2)+n_e\right)+\mathbbm{h}\left(h_e\alpha x_1+n_e\right) \\
\geq&\alpha\mathbbm{h}\left(h_2x_1+n_2\right)+(1-\alpha)\mathbbm{h}\left(h_2x_2+n_2\right)-\mathbbm{h}\left(h_2\alpha x_1+n_2\right) \nonumber \\
&-\mathbbm{h}\left(h_e(\alpha x_1+(1-\alpha)x_2)+n_e\right)+\mathbbm{h}\left(h_e\alpha x_1+n_e\right) \label{eq_r2s_direct_pf_1} \\
\geq&\frac{1}{2}\log\left(e^{2\mathbbm{h}\left(h_2x_1\right)}+e^{2\mathbbm{h}\left(n_2\right)}\right)-\frac{1}{2}\log\left(2\pi e\left(h_2^2\alpha^2\frac{A^2}{3}+1\right)\right) \nonumber \\
&-\frac{1}{2}\log\left(2\pi e\left(h_e^2\alpha^2\frac{A^2}{3}+h_e^2(1-\alpha)^2\frac{A^2}{3}+1\right)\right)+\frac{1}{2}\log\left(e^{2\mathbbm{h}\left(h_e\alpha x_1\right)}+e^{2\mathbbm{h}\left(n_e\right)}\right) \label{eq_r2s_direct_pf_2} \\
\geq&\frac{1}{2}\log\left(e^{2\mathbbm{h}\left(h_2x_1\right)}+e^{2\mathbbm{h}\left(n_2\right)}\right)-\frac{1}{2}\log\left(2\pi e\left(h_2^2\alpha^2\frac{A^2}{3}+1\right)\right) \nonumber \\
&-\frac{1}{2}\log\left(2\pi e\left(h_e^2\frac{A^2}{3}+1\right)\right)+\frac{1}{2}\log\left(e^{2\mathbbm{h}\left(h_e\alpha x_1\right)}+e^{2\mathbbm{h}\left(n_e\right)}\right) \label{eq_r2s_direct_pf_3} \\
=&\frac{1}{2}\log\left(h_2^24A^2+2\pi e\right)-\frac{1}{2}\log\left(2\pi e\left(h_2^2\alpha^2\frac{A^2}{3}+1\right)\right) \nonumber \\
&-\frac{1}{2}\log\left(2\pi e\left(h_e^2\frac{A^2}{3}+1\right)\right)+\frac{1}{2}\log\left(h_2^2\alpha^24A^2+2\pi e\right) \\
=&r_{2,s},
\end{align}
where (\ref{eq_r2s_direct_pf_1}) follows by Jensen's inequality (concavity of differential entropy) \cite{cover}; (\ref{eq_r2s_direct_pf_2}) follows by using EPI to lower bound the positive terms of (\ref{eq_r2s_direct_pf_1}) together with the fact that $h_2x_1+n_2$ and $h_2x_2+n_2$ have the same distribution, and plugging in a Gaussian $x_1$ and $x_2$, instead of uniform, with the same variances to upper bound the negative terms of (\ref{eq_r2s_direct_pf_1}); and (\ref{eq_r2s_direct_pf_3}) follows since $\alpha\leq1$. This concludes the proof.

\subsection{Proof of Theorem~\ref{thm_jam}} \label{apndx_thm_jam}

We first note that, different from direct transmission, over here we have another random variable $z$ involved in the calculations. To emphasize the difference, we denote the secrecy rates in (\ref{eq_c1s}) and (\ref{eq_c2s}) by $c_{1,s}^J$ and $c_{2,s}^J$, respectively. We now proceed with the same approach as that followed in the proof of Theorem~\ref{thm_direct}. Specifically, we let $x_1$ and $x_2$ be two independent uniformly distributed random variables on $[-A_\gamma,A_\gamma]$, and let $z$ be uniformly distributed on $\left[-\bar{A},\bar{A}\right]$, independently of $x_1$ and $x_2$. We then expand the mutual information terms constituting $c_{1,s}^J$ and $c_{2,s}^J$ in terms of differential entropy, lower bound positive terms by EPI (and Jensen's inequality if need be), and upper bound negative terms by plugging in Gaussian random variables with the same variances, instead of uniform. Specific justifications of intermediate steps are as in the proof of Theorem~\ref{thm_direct} and are thus omitted for brevity. We also drop the superscript $+$ for convenience.

A lower bound on $c_{1,s}^J$ is now given by
\begin{align}
c_{1,s}^J=&\mathbbm{I}\left(x_1;h_1\alpha x_1+n_1\right)-\mathbbm{I}\left(x_1;h_e\alpha x_1+{\bm g}_e^T{\bm J}_oz+n_e\right) \\
=&\mathbbm{h}\left(h_1\alpha x_1+n_1\right)-\mathbbm{h}\left(n_1\right)-\mathbbm{h}\left(h_e\alpha x_1+{\bm g}_e^T{\bm J}_oz+n_e\right)+\mathbbm{h}\left({\bm g}_e^T{\bm J}_oz+n_e\right) \\
\geq&\frac{1}{2}\log\left(e^{2\mathbbm{h}\left(h_1\alpha x_1\right)}+e^{2\mathbbm{h}\left(n_1\right)}\right) -\frac{1}{2}\log(2\pi e) \nonumber \\
&-\frac{1}{2}\log\left(2\pi e\left(h_e^2\alpha^2\frac{A^2_\gamma}{3}+\left({\bm g}_e^T{\bm J}_o\right)^2\frac{\bar{A}^2}{3}+1\right)\right)+\frac{1}{2}\log\left(e^{2\mathbbm{h}\left({\bm g}_e^T{\bm J}_oz\right)}+e^{2\mathbbm{h}\left(n_e\right)}\right) \\
=&\frac{1}{2}\log\left(h_1^2\alpha^24A^2_\gamma+2\pi e\right) -\frac{1}{2}\log(2\pi e)\nonumber \\
&-\frac{1}{2}\log\left(2\pi e\left(h_e^2\alpha^2\frac{A^2_\gamma}{3}+\left({\bm g}_e^T{\bm J}_o\right)^2\frac{\bar{A}^2}{3}+1\right)\right)+\frac{1}{2}\log\left(\left({\bm g}_e^T{\bm J}_o\right)^2\alpha^24\bar{A}^2+2\pi e\right) \\
=&r_{1,s}^J.
\end{align}
Similarly, we lower bound $c_{2,s}^J$ as follows:
\begin{align}
c_{2,s}^J=&\mathbbm{I}\left(x_2;h_2(\alpha x_1+(1-\alpha)x_2)+n_2\right)-\mathbbm{I}\left(x_2;h_e(\alpha x_1+(1-\alpha)x_2)+{\bm g}_e^T{\bm J}_oz+n_e\right) \\
=&\mathbbm{h}\left(h_2(\alpha x_1+(1-\alpha)x_2)+n_2\right)-\mathbbm{h}\left(h_2\alpha x_1+n_2\right) \nonumber \\
&-\mathbbm{h}\left(h_e(\alpha x_1+(1-\alpha)x_2)+{\bm g}_e^T{\bm J}_oz+n_e\right)+\mathbbm{h}\left(h_e\alpha x_1+{\bm g}_e^T{\bm J}_oz+n_e\right) \\
\geq&\alpha\mathbbm{h}\left(h_2x_1+n_2\right)+(1-\alpha)\mathbbm{h}\left(h_2x_2+n_2\right)-\mathbbm{h}\left(h_2\alpha x_1+n_2\right) \nonumber \\
&-\mathbbm{h}\left(h_e(\alpha x_1+(1-\alpha)x_2)+{\bm g}_e^T{\bm J}_oz+n_e\right)+\mathbbm{h}\left(h_e\alpha x_1+{\bm g}_e^T{\bm J}_oz+n_e\right) \\
\geq&\frac{1}{2}\log\left(e^{2\mathbbm{h}\left(h_2x_1\right)}+e^{2\mathbbm{h}\left(n_2\right)}\right)-\frac{1}{2}\log\left(2\pi e\left(h_2^2\alpha^2\frac{A^2_\gamma}{3}+1\right)\right) \nonumber \\
&-\frac{1}{2}\log\left(2\pi e\left(h_e^2\alpha^2\frac{A^2_\gamma}{3}+h_e^2(1-\alpha)^2\frac{A^2_\gamma}{3}+\left({\bm g}_e^T{\bm J}_o\right)^2\frac{\bar{A}^2}{3}+1\right)\right) \nonumber\\
&+\frac{1}{2}\log\left(e^{2\mathbbm{h}\left(h_e\alpha x_1\right)}+e^{2\mathbbm{h}\left({\bm g}_e^T{\bm J}_oz\right)}+e^{2\mathbbm{h}\left(n_e\right)}\right) \\
\geq&\frac{1}{2}\log\left(e^{2\mathbbm{h}\left(h_2x_1\right)}+e^{2\mathbbm{h}\left(n_2\right)}\right)-\frac{1}{2}\log\left(2\pi e\left(h_2^2\alpha^2\frac{A^2_\gamma}{3}+1\right)\right) \nonumber \\
&-\frac{1}{2}\log\left(2\pi e\left(h_e^2\frac{A^2_\gamma}{3}+\left({\bm g}_e^T{\bm J}_o\right)^2\frac{\bar{A}^2}{3}+1\right)\right)+\frac{1}{2}\log\left(e^{2\mathbbm{h}\left(h_e\alpha x_1\right)}+e^{2\mathbbm{h}\left({\bm g}_e^T{\bm J}_oz\right)}+e^{2\mathbbm{h}\left(n_e\right)}\right) \\
=&\frac{1}{2}\log\left(h_2^24A^2_\gamma+2\pi e\right)-\frac{1}{2}\log\left(2\pi e\left(h_2^2\alpha^2\frac{A^2_\gamma}{3}+1\right)\right) \nonumber \\
&-\frac{1}{2}\log\left(2\pi e\left(h_e^2\frac{A^2_\gamma}{3}+\left({\bm g}_e^T{\bm J}_o\right)^2\frac{\bar{A}^2}{3}+1\right)\right)+\frac{1}{2}\log\left(h_2^2\alpha^24A^2_\gamma+\left({\bm g}_e^T{\bm J}_o\right)^2\alpha^24\bar{A}^2+2\pi e\right) \\
=&r_{2,s}^J.
\end{align}
This concludes the proof.

\subsection{Proof of Theorem~\ref{thm_dec}} \label{apndx_thm_dec}

We let the relays employ the same decoding technique of the strong user: first decode the weak user's message by treating the strong user's interfering signal as noise, and then use successive interference cancellation to decode the strong user's message. Using the decode-and-froward lower bound in \cite[Theorem 16.2]{elgamalKim}, the following secrecy rates are achievable:
\begin{align}
c_{1,s}^{DF}=&\frac{1}{2}\!\left[\min\!\left\{\!\mathbbm{I}\left(x,x_r;y_1,y_1^r|x_2,\tilde{x}_2\right),\min_i\mathbbm{I}\left(x_1;y_{r,i}|x_2\right)\!\right\}-\mathbbm{I}(x;y_e|x_2)\right]^+, \\
c_{2,s}^{DF}=&\frac{1}{2}\left[\min\left\{\mathbbm{I}\left(x_2,\tilde{x}_2;y_2,y_2^r\right),\min_i\mathbbm{I}\left(x_2;y_{r,i}\right)\right\}-\mathbbm{I}(x_2;y_e)\right]^+,
\end{align}
where the extra $\frac{1}{2}$ term is due to sending the same information over two phases of equal durations. By the independence of $x_j$ and $\tilde{x}_j$, $j=1,2$, we have
\begin{align}
&\mathbbm{I}\left(x_1,\tilde{x}_1;y_1,y_1^r|x_2,\tilde{x}_2\right)=\mathbbm{I}\left(x_1;h_1\alpha x_1+n_1\right)+\mathbbm{I}\left(\tilde{x}_1;{\bm g}_1^T{\bm d}_o\alpha\tilde{x}_1+n_1^r\right), \\
&\mathbbm{I}\left(x_2,\tilde{x}_2;y_2,y_2^r\right)=\mathbbm{I}\left(x_2;h_2\alpha (\alpha x_1+(1-\alpha)x_2)+n_2\right)+\mathbbm{I}\left(\tilde{x}_2;{\bm g}_2^T{\bm d}_o(\alpha\tilde{x}_1+(1-\alpha)\tilde{x}_2)+n_2^r\right).
\end{align}

To derive the lower bounds on $c_{1,s}^{DF}$ and $c_{2,s}^{DF}$, we proceed as in the proof of Theorem~\ref{thm_direct} by lower bounding the positive terms above by EPI (and Jensen's inequality if need be), and upper bounding the negative terms above by plugging in Gaussian random variables with the same variances instead of uniform. This directly gives $r_{1,s}^{DF}$ and $r_{2,s}^{DF}$. Specific details are merely the same as in the proof of Theorem~\ref{thm_direct} and are omitted for brevity.

\subsection{Proof of Theorem~\ref{thm_amp}} \label{apndx_thm_amp}

We note that the $j$th user, $j=1,2$, can view the system as the following $1\times2$ SIMO system:
\begin{align}
\begin{bmatrix}y_j\\y_j^r\end{bmatrix}=\begin{bmatrix}h_j\\{\bm g}_j^T\texttt{diag}\left({\bm h}_r\right){\bm a}_o\end{bmatrix}x+\begin{bmatrix}n_j\\ \tilde{n}_j^r\end{bmatrix},
\end{align}
where the noise term $\tilde{n}_j^r\triangleq{\bm g}_j^T\texttt{diag}\left({\bm n}_r\right){\bm a}_o+n_j^r$, which is $\sim\mathcal{N}\left(0,1+\left({\bm g}_j^T{\bm a}_o\right)^2\right)$. The $j$th user then applies the capacity achieving maximal ratio combining \cite{tse-wireless} to get the following sufficient statistic:
\begin{align}
\tilde{y}_j\triangleq&h_jy_j+\frac{{\bm g}_j^T\texttt{diag}\left({\bm h}_r\right){\bm a}_o}{1+\left({\bm g}_j^T{\bm a}_o\right)^2}y_j^r \\
\triangleq&h_jy_j+\frac{h_{j,r}}{\sigma^2_{j,r}}y_j^r.
\end{align}
Therefore, the following secrecy rates are now achievable:
\begin{align}
c_{1,s}^{AF}=&\frac{1}{2}\left[\mathbbm{I}\left(x;\tilde{y}_1|x_2\right)-\mathbbm{I}(x;y_e|x_2)\right]^+, \label{eq_c1s_a} \\
c_{2,s}^{AF}=&\frac{1}{2}\left[\mathbbm{I}\left(x_2;\tilde{y}_2\right)-\mathbbm{I}(x_2;y_e)\right]^+, \label{eq_c2s_a}
\end{align}
where the extra $\frac{1}{2}$ term is due to sending the same information over two phases of equal durations, as in the decode-and-forward scheme. We now proceed with lower bounding the positive mutual information terms in (\ref{eq_c1s_a}) and (\ref{eq_c2s_a}); the negative terms are handled exactly as in the proof of Theorem~\ref{thm_direct}. For the strong user, we have
\begin{align}
\mathbbm{I}\left(x;\tilde{y}_1|x_2\right)&=\mathbbm{h}\left(\left(h_1^2+\frac{h_{1,r}^2}{\sigma_{1,r}^2}\right)\alpha x_1+h_1n_1+\frac{h_{1,r}}{\sigma_{1,r}^2}\tilde{n}_1^r\right)-\mathbbm{h}\left(h_1n_1+\frac{h_{1,r}}{\sigma_{1,r}^2}\tilde{n}_1^r\right) \\
&\geq\frac{1}{2}\log\left(\!e^{2\mathbbm{h}\left(\left(h_1^2+\frac{h_{1,r}^2}{\sigma_{1,r}^2}\right)\alpha x_1\right)}+e^{2\mathbbm{h}\left(h_1n_1\right)}+e^{2\mathbbm{h}\left(\frac{h_{1,r}}{\sigma_{1,r}^2}\tilde{n}_1^r\right)}\!\right)\!-\!\frac{1}{2}\log\left(\!(2\pi e)\left(h_1^2+\frac{h_{1,r}^2}{\sigma_{1,r}^2}\right)\!\right) \\
&=\frac{1}{2}\log\left(\!\left(h_1^2+\frac{h_{1,r}^2}{\sigma_{1,r}^2}\right)^2\alpha^24A^2_\gamma+(2\pi e)\left(h_1^2+\frac{h_{1,r}^2}{\sigma_{1,r}^2}\right)\!\right)\!-\!\frac{1}{2}\log\left(\!(2\pi e)\left(h_1^2+\frac{h_{1,r}^2}{\sigma_{1,r}^2}\right)\!\right) \\
&=\frac{1}{2}\log\left(1+\frac{2\kappa_1^2\alpha^2A^2_\gamma}{\pi e}\right).
\end{align}
Similarly, for the weak user, we have
\begin{align}
\mathbbm{I}\left(x_2;\tilde{y}_2\right)&=\mathbbm{h}\left(\left(h_2^2+\frac{h_{2,r}^2}{\sigma_{2,r}^2}\right)\left(\alpha x_1+(1-\alpha)x_2\right)+h_2n_2+\frac{h_{2,r}}{\sigma_{2,r}^2}\tilde{n}_2^r\right) \nonumber \\
&\hspace{.2in}-\mathbbm{h}\left(\left(h_2^2+\frac{h_{2,r}^2}{\sigma_{2,r}^2}\right)\alpha x_1+h_2n_2+\frac{h_{2,r}}{\sigma_{2,r}^2}\tilde{n}_2^r\right) \\
&\geq\alpha\mathbbm{h}\!\left(\!\left(\!h_2^2+\frac{h_{2,r}^2}{\sigma_{2,r}^2}\!\right)x_1+h_2n_2+\frac{h_{2,r}}{\sigma_{2,r}^2}\tilde{n}_2^r\!\right)\!+\!(1\!-\!\alpha)\mathbbm{h}\!\left(\!\left(\!h_2^2+\frac{h_{2,r}^2}{\sigma_{2,r}^2}\!\right)x_2+h_2n_2+\frac{h_{2,r}}{\sigma_{2,r}^2}\tilde{n}_2^r\!\right) \nonumber \\
&\hspace{.2in}-\frac{1}{2}\log\left((2\pi e)\left(\left(h_2^2+\frac{h_{2,r}^2}{\sigma_{2,r}^2}\right)^2\frac{A^2_\gamma}{3}+h_2^2+\frac{h_{2,r}^2}{\sigma_{2,r}^2}\right)\right) \\
&\geq\frac{1}{2}\log\left(e^{2\mathbbm{h}\left(\left(h_2^2+\frac{h_{2,r}^2}{\sigma_{2,r}^2}\right)x_1\right)}+e^{2\mathbbm{h}\left(h_2n_2\right)}+e^{2\mathbbm{h}\left(\frac{h_{2,r}}{\sigma_{2,r}^2}\tilde{n}_2^r\right)}\right) \nonumber \\
&\hspace{.2in}-\frac{1}{2}\log\left((2\pi e)\left(\left(h_2^2+\frac{h_{2,r}^2}{\sigma_{2,r}^2}\right)^2\frac{A^2_\gamma}{3}+h_2^2+\frac{h_{2,r}^2}{\sigma_{2,r}^2}\right)\right) \\
&=\frac{1}{2}\log\left(\left(h_2^2+\frac{h_{2,r}^2}{\sigma_{2,r}^2}\right)^24A^2_\gamma +(2\pi e)\left(h_2^2+\frac{h_{2,r}^2}{\sigma_{2,r}^2}\right)\right) \nonumber \\
&\hspace{.2in}-\frac{1}{2}\log\left((2\pi e)\left(\left(h_2^2+\frac{h_{2,r}^2}{\sigma_{2,r}^2}\right)^2\frac{A^2_\gamma}{3}+h_2^2+\frac{h_{2,r}^2}{\sigma_{2,r}^2}\right)\right) \\
&=\frac{1}{2}\log\left(\frac{1+\frac{2\kappa_2^2A^2_\gamma}{\pi e}}{1+\frac{\kappa_2^2\alpha^2A^2_\gamma}{3}}\right).
\end{align}
This concludes the proof.



\end{document}